\documentclass[aps,twocolumn,10pt,floatfix, superscriptaddress,longbibliography,pra]{revtex4-1}
\usepackage{graphicx,amsmath}
\usepackage{color}
\usepackage{natbib}
\usepackage{ulem}

\setcitestyle{authoryear,round}
\usepackage{siunitx}
\newcommand\Rey{\mbox{\textit{Re$_s$}}}  
\newcommand\We{\mbox{\textit{We}}}  
\newcommand\Fr{\mbox{\textit{Fr$_b$}}}  
\newcommand\Frcent{\mbox{\textit{Fr$_\text{cent}$}}}  
\newcommand{\perc}[1]{ \SI{#1}{\percent} }
\newcommand{\e}[1]{\times 10^{#1}}

\newcommand{\TCf}{Taylor--Couette }

\begin{document}

\title{Effect of axially varying sandpaper roughness on bubbly drag reduction in Taylor--Couette turbulence}
\author{Pim A. Bullee}
\affiliation{Physics of Fluids, Max Planck centre Twente for Complex Fluid Dynamics, MESA+ Research Institute and J. M. Burgers Centre for Fluid Dynamics, University of Twente, P.O. Box 217, 7500 AE Enschede, The Netherlands}
\affiliation{Soft matter, Fluidics and Interfaces, MESA+ Research Institute, University of Twente, P.O. Box 217, 7500 AE Enschede, The Netherlands}
\author{Dennis Bakhuis}
\author{Rodrigo Ezeta}
\author{Sander G. Huisman}
\email{s.g.huisman@utwente.nl}
\affiliation{Physics of Fluids, Max Planck centre Twente for Complex Fluid Dynamics, MESA+ Research Institute and J. M. Burgers Centre for Fluid Dynamics, University of Twente, P.O. Box 217, 7500 AE Enschede, The Netherlands}
\author{Chao Sun}
\email{chaosun@tsingua.edu.cn}
\affiliation{Centre for Combustion Energy, Key Laboratory for Thermal Science and Power Engineering of Ministry of Education, Department of Energy and Power Engineering, Tsinghua University, Beijing 100084, China}
\affiliation{Department of Engineering Mechanics, School of Aerospace Engineering, Tsinghua University, Beijing 100084, China }
\author{Rob G. H. Lammertink}
\affiliation{Soft matter, Fluidics and Interfaces, MESA+ Research Institute, University of Twente, P.O. Box 217, 7500 AE Enschede, The Netherlands}
\author{Detlef Lohse}
\email{d.lohse@utwente.nl}
\affiliation{Physics of Fluids, Max Planck centre Twente for Complex Fluid Dynamics, MESA+ Research Institute and J. M. Burgers Centre for Fluid Dynamics, University of Twente, P.O. Box 217, 7500 AE Enschede, The Netherlands}
\affiliation{Max Planck Institute for Dynamics and Self-Organization, Am Fassberg 17, 37077 G\"{o}ttingen, Germany}

\date{\today}

\begin{abstract}
We experimentally investigate the influence of alternating rough and smooth walls on bubbly drag reduction (DR). To this end, we apply rough sandpaper bands of width $s$ between $\SI{48.4}{mm}$ and $\SI{148.5}{mm}$, and roughness height $k = \SI{695}{\um}$, around the smooth inner cylinder of the Twente Turbulent \TCf facility. Between two sandpaper bands, the inner cylinder is left uncovered over similar width $s$, resulting in alternating rough and smooth bands, forming a constant pattern in axial direction. We measure the DR in water that originates from introducing air bubbles to the fluid at (shear) Reynolds numbers $\Rey$ ranging from $0.5\e{6}$ to $1.8\e{6}$. Results are compared to bubbly DR measurements with a completely smooth inner cylinder and an inner cylinder that is completely covered with sandpaper of the same roughness $k$. The outer cylinder is left smooth for all variations. The results are also compared to bubbly DR measurements where a smooth outer cylinder is rotating in opposite direction to the smooth inner cylinder. This counter rotation induces secondary flow structures that are very similar to those observed when the inner cylinder is composed of alternating rough and smooth bands. For the measurements with roughness, the bubbly DR is found to initially increase more strongly with $\Rey$, before levelling off to reach a value that no longer depends on $\Rey$. This is attributed to a more even axial distribution of the air bubbles, resulting from the increased turbulence intensity of the flow compared to flow over a completely smooth wall at the same $\Rey$. The air bubbles are seen to accumulate at the rough wall sections in the flow. Here, locally, the drag is largest and so the drag reducing effect of the bubbles is felt strongest. Therefore, a larger maximum value of bubbly DR is found for the alternating rough and smooth walls compared to the completely rough wall.
\end{abstract}

\maketitle

\noindent{\bf Highlights}
\begin{itemize}
\item Study on the the effect of roughness on bubbly drag reduction in Taylor--Couette flow.
\item Bubbles accumulate on the roughness, instead of being trapped in the turbulent Taylor vortices.
\item Roughness leads to a larger effect of bubbly drag reduction.
\end{itemize}
\newpage


\clearpage
\section{Introduction}
Wall-bounden high Reynolds number flows are known to experience a significant increase in drag due to roughness~\citep{Jimenez2004,Marusic2010,Flack2010,Flack2014}. To reduce energy costs, the aim is to reduce this frictional resistance. Therefore drag reduction (DR) in wall-bounded turbulent flows using bubble injection has been a matter of study for long time~\citep{Ceccio2010,Murai2014,Verschoof2018}. Promising applications can be found in the maritime industry, where a reduction of the ship drag force will result in reduced fuel consumption. The total drag of a ship is composed of form drag (related to the design of the hull) and skin friction drag, of which the latter is dependent on the surface properties of the hull and increases drastically with biofouling growth~\citep{Jimenez2004,Schultz2007,Flack2010,Flack2014}. While air bubble DR is commonly studied in laboratory set~ups that make use of smooth walls, we study air bubble DR in turbulent flows over heterogeneous rough walls. The present investigation is aimed at gaining better understanding of the mechanism of the bubbly DR, contributing to the research of its industrial applications..

\subsection{Taylor--Couette}
The flow geometry we use to study bubble DR over rough walls is the Taylor--Couette (TC) geometry. The flow is generated between two concentric, independently-rotating cylinders. The radii of the inner and outer cylinder are given by $r_i$ and $r_o$, respectively, and the width of the gap between the cylinders is $d = r_o - r_i$, see also figure~\ref{fig:setup}. Together with the height of the cylinders $L$, two geometrical parameters can be defined: the radius ratio $\eta = r_i/r_o$, and the aspect ratio $\Gamma = L/d$. The Taylor--Couette flow is rich in flow structures~\citep{Andereck1986}, and is used in many fundamental studies, such as magnetohydrodynamics~\citep{Chandrasekhar1961,Balbus1991}, astrophysics~\citep{Richard1999}, hydrodynamic stability analysis~\citep{Taylor1923}, and drag reduction~\citep{Srinivasan2015}. Apart from the fundamental knowledge gained from the geometry, it has also lend itself for a wide variety of applications in the fields of multi-phase flow and boiling~\citep{Ezeta2019}, medical engineering~\citep{Beaudoin1989,Wereley1999}, turbo-machinery~\citep{Jeng2007}, and beyond. The TC geometry is a mathematically well defined and closed system, with a well defined energy balance~\citep{Eckhardt2007}, making it one of the canonical systems to study the physics of fluids. See the reviews by~\cite{Fardin2014,Grossmann2016} for a broader introduction to, and an overview of different studies, on TC flow. The working fluid between the cylinders is set in motion by the rotation of either one or both cylinders, which generates a shear flow. We define a characteristic shear Reynolds number $\Rey$ using the different geometric parameters of the system, the properties of the fluid and the rotation rates of both cylinders as
    \begin{equation}\label{eq:Rey}
    \Rey = \frac{r_i (\omega_i - \omega_o) d}{\nu}.
    \end{equation}
Here $\omega_{i,o}$ denotes the rotation rates of the inner (subscript $i$) and outer (subscript $o$) cylinder. The fluid kinematic viscosity is denoted by $\nu$.

In TC flow, angular momentum is transported from the inner- to the outer cylinder. The transport of this angular momentum is linearly related to the torque, which is needed to keep the cylinders spinning at constant angular velocity, and hence to the energy input to the system. The torque can be measured with relative  ease and accuracy, making the TC system very well suited to measure fluid drag. Due to its geometry, stable secondary flow structures are formed, in the form of rolls, in the gap between the inner- and outer cylinder. Even when the flow is highly turbulent, some form of order can be discovered, when radially outward transported fluid forms organised structures with radially inward moving fluid in the form of so-called turbulent Taylor vortices~\citep{Andereck1986}. For high $\Rey$ this only occurs in the counter-rotating regime, when the outer cylinder is rotating in opposite direction to the inner cylinder~\citep{Ostilla-Monico2014,Grossmann2016}. These rolls enhance the angular momentum transport, while their strength varies with both $\Rey$ and the negative rotation ratio between inner- and outer cylinder $a$ defined as
    \begin{align}\label{eq:a}
        a = -\frac{\omega_o}{\omega_i}.
    \end{align}
When $a = a_\text{optimal}$, the angular momentum transport is the highest, and the vortices are the strongest, which leads to the largest value of the drag on the cylinders for that specific shear Reynolds number. At $\eta=0.716$, $a_\text{optimal} = 0.36$ for $\Rey = \mathcal{O}(10^6)$~\citep{Huisman2014}, though the dependence on $\eta$ is complicated~\citep{Ostilla2014}.
Also larger fluctuations of the fluid velocity within the gap are found in the counter-rotating regime compared with those observed when the outer-cylinder is stationary~\citep{Dong2008,Huisman2013b}.

\subsection{Bubbly drag reduction}
An overview of different studies on air bubble injection DR is given in the review articles by \cite{Ceccio2010} and \cite{Murai2014}. For bubbly DR to be effective, the injected air bubbles need to stay close to the wall~\citep{vandenBerg2007}. When the bubbles migrate away from the wall, the DR effect will be lost~\citep{Watanabe1998,Lu2005,Murai2005,Sanders2006,Elbing2008,Lu2008,Murai2014}. To achieve high values of bubbly DR, bubbles also need to have a large Weber number~\citep{vanGils2013,Verschoof2016,Spandan2018}. Following the definition of \citep{vandenBerg2005} we define the Weber number $\text{We}=\rho u' D / \sigma$, where $\rho$ is the density of liquid, $u'$ is the standard deviation of the fluid velocity fluctuations, $D$ the bubble diameter, and $\sigma$ the surface tension at the bubble-liquid interface. When the Weber number is small ($\We \ll 1$), bubbles are more easily transported by the turbulent flow, moving away from the boundaries. The amount of bubbly DR scales linearly with the amount of injected air in an open system~\citep{Elbing2008}. When the amount of bubbles near the surface is sufficiently large, an air layer is formed~\citep{Zverkhovskyi2014,Rotte2016}.

In Taylor--Couette turbulence it was found that wall roughness rib-like elements induce strong secondary flows, that transported the bubbles away from the wall and decreased the DR~\citep{vandenBerg2007,Verschoof2018}. This might be an effect related to the flow geometry, since in channel flow, increased microbubble DR was found for turbulent flow over sandpaper rough walls compared to smooth walls~\citep{Deutsch2004}. However, it could also be attributed to a larger baseline drag for the rough walls. The relation between DR and gas injection rate was very similar for all rough and smooth cases~\citep{Deutsch2004}. In the limit of high gas injection rates, bubbly DR turns into to gas (air) layer DR as described in~\cite{Elbing2008}. The excess gas no longer forms bubbles, but instead a thin sheet is formed, decoupling the wall from the working liquid. Typical values of DR observed in this regime are \SI{90 +- 10}{\percent}~\citep{Elbing2008}. Similar values of DR were observed by \cite{Saranadhi2016}, from vapour bubbles created at the inner cylinder~\citep{Saranadhi2016}. When an air layer is formed, a further increase of air injection rate does not further decrease the drag~\citep{Elbing2008}, a limit also reached by~\cite{Deutsch2004} for all rough and smooth cases. Where the majority of (fundamental) studies of bubbly DR make use of smooth walls, it also very relevant to include rough walls in these studies, as in applications surfaces typically feature some kind of roughness.

\subsection{Spanwise-varying roughness}
Especially when the Reynolds numbers are large and the flows become turbulent, even small (\si{\micro \meter}-scale) roughness elements are felt by the flow. Hence, in practice, most surfaces are rough, or at least feature roughness to some degree. Although the \si{\micro \meter}-scale roughness elements might seem too small to be of influence on the flow, compared to the smallest length scales found in such a turbulent flow, they are very relevant. As a result, turbulent flows over rough walls are extensively studied. For a complete overview of the influence of wall roughness, we refer to the reviews and work by~\cite{Jimenez2004}~and~\cite{Flack2010, Flack2014}. The majority of the studies focus, however, on homogeneous roughness, with a typical roughness feature size $k$ much smaller than the major length scale of the flow $d$, e.g. the (half) height for channel-flow systems, though exceptions with large $k$ exist \citep{amir2014,mazzuoli2017}. However, in practice not all surfaces have homogeneous roughness, but rather roughness of a distributed and heterogeneously rough kind. Examples of heterogeneous obstacle roughness include connections and fasteners (welding seams, pipe joints), and damages of a larger length scale, but also atmospheric flows over a varying terrain of grass and woodlands~\citep{Ren2011}. On ship hulls in the maritime industry, examples also include clusters of biofouling (barnacles). For roughness variations of a smaller length scale, examples found in many industrial applications include corrosion, micro-fouling (bio-slime), and variations in coating condition~\citep{Yeginbayeva2018}. Therefore it is industrially very relevant to include surfaces of non-homogeneous roughness in studies on roughness, and also to consider their influence on bubbly DR.

An important parameter in quantifying heterogeneous spanwise-varying rough surfaces is the size of the alternating rough and smooth patches, $L_r$ and $L_s$. When the rough and smooth patches are of equal size, so when $L_r = L_s$, typically a single parameter $s$ is used for the patch size. The flow over a rough section will experience a higher wall shear stress compared to the flow over a smooth section. Streamwise roll motions are induced at the edges between rough and smooth patches where sharp stresses are also observed~\citep{Hinze1967,Barros2014,Willingham2014,Chung2018}.
Instead, for very small and very large patch spacings (e.g.~$s/d <0.39$ or $s/d > 6.28$, where $d$ is the half-channel height), the induced secondary flows are either not strong and large enough, or not able to interact, thereby havin a lesser  effect on the bulk flow~\citep{Chung2018}. Between these extremes, the roll motions were seen to interact with each other, generating a wall-normal velocity that does influence the bulk flow, breaking with Townsend's hypothesis \citep{Townsend1976} of outer layer similarity, that states that the turbulent flow in the bulk region is determined by the wall shear stress only~\citep{Chung2018}.

In Taylor--Couette flow, the effects of span-wise varying roughness on the flow was studied both experimentally and numerically by~\cite{Bakhuis2019}. The rough patches on the inner cylinder consisted of sandpaper bands, resulting in an axially varying pattern of rough and smooth bands. As the flow was driven by the rotation of the inner cylinder, larger velocities and turbulent fluctuations were found in the flow near the rough patches, compared to the smooth patches~\citep{Bakhuis2019}. Note that this is different from pressure driven flows, such as channel or pipe flow, where a lower velocity will be found in the flow near the rough patches, compared to the smooth patches. The velocity differences triggered the formation of secondary flow structures in the form of rolls, that are similar to the turbulent Taylor vortices found for smooth wall Taylor--Couette flow for $a>0$. By changing the size (axial height) of the smooth and rough bands, the sizes of the rolls were manipulated, as the (radially) outward flow near the rough patch forms pairs with the inward flows near the two adjacent smooth patches (axially above and below the rough patch), resulting in the formation of roll structures~\citep{Bakhuis2019}.

In general, secondary flows will increase the momentum transfer, as advection is more effective in this than diffusion. Therefore, the drag will also increase. The momentum and drag increase due to alternating boundary conditions was studied in different flow configurations, for instance in Rayleigh--B\'{e}nard flow~\citep{Bakhuis2018} (heat-transfer rather than momentum transfer), in pipe flow~\citep{Chan2018}, channel flow~\citep{Chung2018}, and Taylor--Couette flow~\citep{vanGils2012}.

\subsection{Bubble position in the flow}
To achieve air bubbly DR, the distribution of the bubble positions in the domain is important~\citep{Fokoua2015}. The dynamics and kinetics of the bubbles in a turbulent flow are very complex, and experimentally obtaining them is even more complicated as the length and time scales are small and even for a small void fraction (say $1\%$) other bubbles occlude the view.

Excellent reviews of the dynamics of bubbles and studies related to this topic are given in the works of e.g.~\cite{Magnaudet2000}~and~\cite{Lohse2018}. Generally, bubbles in turbulent flow are observed to cluster in regions of high vorticity and low pressure~\citep{Mazzitelli2003,Climent2007}.

In Taylor--Couette turbulence, the bubble position depends on the interplay between buoyancy force, the centripetal forces of both the mean flow displacement (rotation of the inner cylinder) and the Taylor vortices as well as the action of smaller turbulent structures~\citep{Djeridi1999,Chouippe2014,Fokoua2015,Lohse2018}. The central control parameter in this is the bubble Froude number, defined as the ratio between centrifugal and gravitational forces acting on the bubble
\begin{equation}\label{eq:Fr}
    \Fr = \frac{r_i \omega_i}{\sqrt{g r_b }},
\end{equation}
with $r_b$ the bubble radius and $g$ the gravitational acceleration. For small Froude numbers $\Fr < 1$, buoyancy effects are dominating. This typically occurs at low Reynolds numbers, when $\omega_i$ is small, and the strength of the turbulent Taylor vortices is only marginal, or  when bubbles are large~\citep{Climent2007,Lohse2018}. Here the mechanism for DR are associated with the rising bubbles that destroy the Taylor vortices, reducing the transport of angular momentum from inner to outer cylinder~\citep{Spandan2018,Lohse2018}.

When $Fr_b$ is large enough, the stronger Taylor vortices that form at larger Reynolds numbers can, trap the bubbles near their cores, and also at outflow regions close to the inner cylinder~\citep{Climent2007,Fokoua2015}. When the bubbles are trapped and passively advected by the Taylor vortices, i.e., for large enough $\Fr$, their influence on the global drag is minimal~\citep{Lohse2018,Spandan2018}. With further increasing Reynolds numbers, the flow dominance of turbulent Taylor vortices decreases~\citep{vanGils2012,Huisman2014} and we might expect the centripetal force from the mean flow displacement to push bubbles towards the inner cylinder. Contextually, smaller turbulent structures tend to disperse the smaller bubbles~\citep{vanGils2013,Chouippe2014}. In the high Reynolds number regime of ultimate turbulence where we operate, stable and unstable roll structures do however still persist in the form of turbulent Taylor vortices~\citep{Huisman2014}.

In order to quantify the balance between the turbulent pressure fluctuations, that distribute the bubbles away from the wall, and the centripetal forces, that push the particle towards the inner-cylinder, \cite{vanGils2013} defined a centripetal Froude number
\begin{equation}\label{eq:Fr_cent}
    \Frcent(r) = \frac{\sigma(u_\theta)^2 / 2r_b}{\langle u_\theta \rangle^2 / r}.
\end{equation}
In this equation, $\sigma(u_\theta)$ is the standard deviation of the azimuthal velocity $u_\theta$, which is related to the turbulent pressure fluctuations at the distance $r-r_i$ from the surface of the inner cylinder.

In this study we build on the work of~\cite{Bakhuis2019}, who have used spanwise-varying roughness to control the secondary flow configurations that show up as turbulent Taylor vortices in high Reynolds number Taylor--Couette flow. In the current work we will use the different secondary flow configurations to study the influence on bubbly DR and the position of bubbles in the flow. With this we provide insight into the mechanisms involving bubbly  DR in high Reynolds number flows. This is relevant for flows over rough, smooth and heterogeneous rough surfaces, giving guidelines to industry for bubbly DR opportunities for a variety of surfaces. \\

\section{Methods}

\setlength{\tabcolsep}{0.7em}
\begin{table}[]
    \centering

    \begin{tabular}{|c|c|c|c|}
        \hline
       $\Rey$~$[10^6]$  & $a$ & Condition IC & $\alpha$  \\
       \hline
         0.5--1.8 & 0 & $\tilde s = 0.61$ & $0\%$ and $2\%$ \\
         0.5--1.8 & 0 & $\tilde s = 0.93$ & $0\%$ and $2\%$ \\
         0.5--1.8 & 0 & $\tilde s = 1.23$ & $0\%$ and $2\%$ \\
         0.5--1.8 & 0 & $\tilde s = 1.87$ & $0\%$ and $2\%$ \\
         0.5--1.8 & 0 & entirely rough & $0\%$ and $2\%$ \\
         0.5--1.8 & 0 & entirely smooth & $0\%$ and $2\%$ \\
         \hline
         0.8 & 0.0--1.0 & entirely smooth & $0\%$ and $2\%$ \\
         1.2 & 0.0--1.0 & entirely smooth & $0\%$ and $2\%$ \\
         1.6 & 0.0--1.0 & entirely smooth & $0\%$ and $2\%$ \\
        \hline
    \end{tabular}

    \caption{Parameters and settings used for the experiments. $a=-\omega_o/\omega_i$ is the rotation ratio, the condition of the inner cylinder (IC) is given, while the outer cylinder is kept smooth, $\tilde s$ is the dimensionless patch size, and $\alpha$ the void fraction of air. Range of values (indicated by --) mean that either the rotation rates or the rotation ratio is changed quasi-statically during the experiment.}
    \label{tab:parameters}
\end{table}

\begin{figure}[ht]
\centering
\includegraphics[scale=1.2]{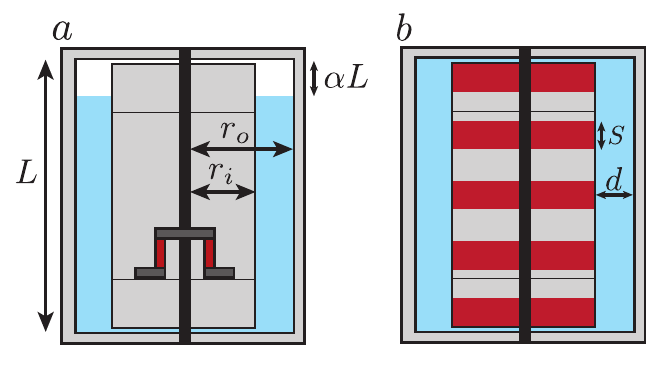}
\caption{Schematic overview of the measurement set up, showing the outer cylinder, three-section inner cylinder and the driving shaft. In a), a smooth inner cylinder is shown, together with the torque sensor that is placed inside the middle section of the inner cylinder and makes the connection to the driving shaft. Also depicted in a) is the void fraction $\alpha$, defining the amount of free air in the system. Out of this air pocket, when the inner cylinder is rotating strongly enough, a two-phase bubbly flow is eventually formed as the result from turbulent mixing. Figure b) shows the inner cylinder coated with sandpaper bands, creating an axially alternating rough/smooth surface. Four different band widths~$s$ are used. Normalized by the gap width $d$ these give values of $\tilde{s} = s/d =$ 1.87, 1.23, 0.93 and 0.61. Since the roughness coverage of the surface is kept constant at $56\%$, we create patterns of 4, 5, 6, and 10 roughness bands, respectively.}\label{fig:setup}
\end{figure}

All experiments were performed in the Twente Turbulent Taylor--Couette facility (T$^3$C) as introduced in~\cite{VanGils2011} and shown schematically in figure~\ref{fig:setup}. The set~up consist of two concentric cylinders of height $L = \SI{927}{mm}$ and radii $r_o = \SI{279.4}{mm}$ and $r_i = \SI{200}{mm}$, resulting in a gap of width $d = r_o - r_i = \SI{79.4}{mm}$. This gives a radius ratio $\eta = {0.716}$, and an aspect ratio $\Gamma = 11.68$. The resulting gap has a volume of $\SI{111}{\liter}$ and is filled with water while leaving out a void fraction $\alpha = (2 \pm 0.2)\,\%$ for air to form bubbles when the working fluid is set in motion. When we study single-phase flow, no air bubbles are introduced to the working fluid and $\alpha = 0$. The range of $\Fr$ studied is approximately $\numrange{30}{500}$, where we base our approximation of the bubble size on~\cite{vanGils2013}.

We vary the rotational frequency of the inner cylinder between \SI{5}{Hz} and \SI{18}{Hz}, whilst keeping the outer cylinder stationary. For this $a = 0$ case, the shear Reynolds number $\Rey$ then ranges from $0.5\e6$ to $1.8\e6$. We also study the influence of the large roll structures that originate from outer cylinder counter-rotation ($a > 0$) on bubbly DR. For this we use three different shear Reynolds numbers: $\Rey = 0.8\e{6}$, $\Rey = 1.2\e{6}$, and $\Rey = 1.6\e{6}$. The ratio of the rotation rates $a$ is then varied between 0 and 1.

The inner cylinder is fabricated from stainless steel and machined in azimuthal direction such that the largest surface roughness is in axial direction and has a value of $k_\text{ic} = \SI{1.6}{\um}$. Using normalization with the length scale of the viscous sublayer $\delta_\nu$, this can be expressed as a maximum roughness of $k_\text{ic} / \delta_\nu = k^+_\text{ic} \approx 1.0$ in wall normal units, reached at the maximum shear Reynolds number. From this it is concluded that the surface of the inner cylinder can be considered to be hydrodynamically smooth in our measurement range~\citep{Schlichting}. The outer cylinder is fabricated from transparent polished PMMA, allowing for optical accessibility of the flow in the gap.

\subsection{Torque measurements}
The amount of energy required to drive the system at set rotational frequencies is determined by the torque measured between the drive shaft and the inner cylinder using a Honeywell 2404-1k hollow reaction torque sensor, and the rotation rate is measured using a magnetic angular encoder. The torque is only measured on the middle part of the three-section inner cylinder ($L_\text{mid} = \SI{536}{mm}$), not to account for end-plate influences that modify the flow near the top and bottom of the system. For our drag measurements, we continuously measure the torque while accelerating the inner cylinder from \SI{5}{Hz} to \SI{18}{Hz} over a period of \SI{78}{\min}. For the measurements with outer cylinder counter-rotation, we start at $a=0$ and increase this to $a=1$ at fixed shear Reynolds number, see Table~\ref{tab:parameters}. The acceleration of the cylinders is the same as for the measurements with a stationary outer cylinder. From the torque $\mathcal{T}$ we calculate the skin friction coefficient $C_f$ defined as
\begin{equation}
    C_f = \frac{\mathcal{T}}{L_{\text{mid}}\rho\nu^2\Rey^2} \:=\: \frac{\mathcal{T}}{L_{\text{mid}}\rho r_i^2 (\omega_i - \omega_o)^2 d^2},
\end{equation}
where $\rho$ and $\nu$ are the density and kinematic viscosity, respectively, of the liquid, and $\Rey$ as defined in equation~\ref{eq:Rey}.
The temperature of the working fluid is continuously measured using a PT100 temperature sensor placed inside the inner cylinder. The density and viscosity of the liquid are temperature corrected using these measurements. To limit the temperature changes of the working liquid that are the result of viscous dissipation inside the liquid, cooling is applied through the top and bottom plate of the set~up, controlling the temperature at \SI{21 +- 0.5}{\celsius}.

\subsection{Axially varying roughness}
By making use of sandpaper belts that are attached to the inner cylinder, we changed the roughness (pattern) of the inner cylinder wall. This is done through the same method and using the same materials as in our previous work~\citep{Bakhuis2019}. Apart from a completely smooth inner cylinder (no sandpaper attached) and a completely rough inner cylinder (whole surface covered with sandpaper) we study different repeating patterns of alternating rough and smooth bands, as shown schematically in figure~\ref{fig:setup}. By using bands of different widths $s$ ($\SI{48.4}{mm}$, $\SI{73,8}{mm}$, $\SI{97.7}{mm}$, and $\SI{148.5}{mm}$), four different patterns were formed of 10, 6, 5, and 4 roughness bands respectively. For each pattern the coverage of the surface with roughness was \SI{56}{\percent}. This is the case for each section of the three-piece inner cylinder (see Fig.~\ref{fig:setup}), and for the cylinder as a whole as well. We can normalize the roughness band width $s$ by the gap width $d$, resulting in values of $\tilde{s} = s/d$ of 0.61, 0.93, 1.23, and 1.87. Together with a completely smooth and completely rough inner cylinder, this results in a total of 6 different variations of the inner cylinder roughness that we studied. The outer cylinder was kept smooth at all times.

\begin{figure*}
    \centering
    \includegraphics[width=0.9\textwidth]{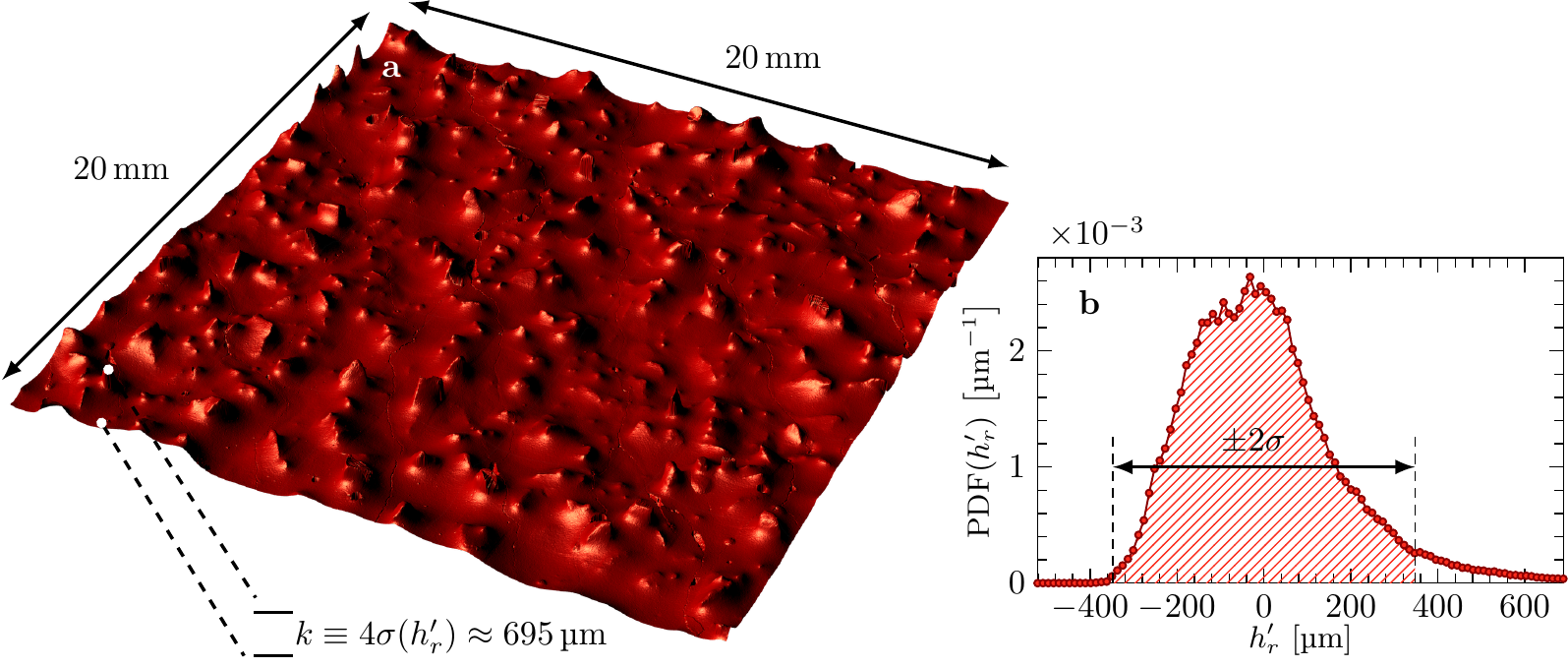}
    \caption{a) Confocal microscopy image of a sample of the sandpaper roughness using the data from~\cite{Bakhuis2019}. b) Height distribution (PDF) of the surface determined using the data from~\cite{Bakhuis2019}. The roughness is determined as $k \approx \SI{695}{\micro\metre}$ using the peak-to-through distance, with the peak and through at $+2\sigma$ and $-2\sigma$ from the mean as described in~\citep{Bakhuis2019}. Figure is adapted from~\citep{Bakhuis2019}.}
    \label{fig:roughness_height}
\end{figure*}

To create the regions of roughness on the inner cylinder, commercially available P36 industrial grade sandpaper belts (VSM XK885Y ceramics plus) were applied using double-sided adhesive tape (Tesa 51970), that together form a \SI{2.5}{\milli\metre} layer. This slight protrusion of the roughness as compared to the smooth regions can have an influence on the direction of the secondary flows as opposing results were found by~\cite{Mejia2013} and~\cite{Vanderwel2015} for recessing and protruding roughness. Compared to the sandpaper width $s$ or the gap width $d$, the protrusion of the roughness is always less than $\SI{5}{\percent}$, which is the smallest protrusion studied by~\cite{Vanderwel2015}. Since our flow is driven by the walls, as opposed to the aforementioned results of channel flow, the direction of the induced secondary flow might again change. A \SI{20}{mm} by \SI{20}{mm} sample of the same sandpaper had been characterised using confocal microscopy with a resolution of \SI{2.5}{\um}~\citep{Bakhuis2019}, see figure~\ref{fig:roughness_height}. With most of the roughness height $h'_r$ within $\pm 2\sigma$ of the mean, the roughness of the surface is defined as the peak-to-valley distance $k \equiv 4\sigma(h_r) \approx \SI{695}{\um}$. In wall normal units this corresponds to a value of $k^+ \approx 434$ for the largest $\Rey$ of $1.8\e6$ and $k^+ \approx 122$ for the smallest $\Rey$ of $5.0\e5$. Hence, all experiments are in the fully rough regime, since over the whole range of $\Rey$ $k^+ > 70$~\citep{Schlichting}. The driving of the flow over the roughness is dominated by pressure forces, whereas on the smooth parts this is purely driven by viscous forces~\citep{Zhu2017,Zhu2018}. On the hull of a ship, a roughness of $k^+ = 122$ would translate to a roughness $k \approx \SI{3}{\mm}$, derived using a flat plate approximation for a $\SI{100}{\metre}$ vessel with a velocity of \SI{10}{m/s}. The typical size of small barnacle biofouling
that grows on underwater ship hulls is about \SI{2.5}{\mm}~\citep{Schultz2004,Demirel2017,Demirel20172}. Our largest roughness $k^+ = 434$ would correspond to $k \approx \SI{13}{\mm}$, following the same flat plate approximation, which is similar to the size of very large barnacles with a typical size of $\SI{10}{\mm}$~\citep{Schultz2004,Demirel2017,Demirel20172}.

\subsection{Counter rotating outer cylinder}
To generate a flow with turbulent Taylor vortices that is similar to the flow encountered in the measurements with an axially varying rough inner cylinder, rotation of the outer cylinder was introduced to the flow over a smooth inner cylinder, in opposite direction of rotation~\citep{Huisman2014}. Since now an additional parameter $a \in [0,1]$ is added to the phase space, we choose to limit ourselves to three different shear Reynolds numbers: $\Rey = 0.8\e{6}$, $\Rey = 1.2\e{6}$, and $\Rey = 1.6\e{6}$. For a fixed $\Rey$, we quasistatically ramp up from $a = 0$ to $a=1$ and measure the torque. The skin friction coefficient $C_f$ is compared between a two-phase flow ($\alpha =\perc{2}$) and a single-phase flow ($\alpha=\perc{0}$).

\subsection{Flow visualizations}
For the flow visualizations a Nikon D800E camera was used with a Sigma \SI{50}{mm} objective. All visualizations were done under the same flow conditions of $\Rey = 0.8 \e{6}$ and $\alpha = \SI{1}{\percent}$. Because of the transparant outer cylinder, the bubbles and the roughness patches can easily be observed by eye.

\section{Results}

\subsection{Flow visualizations}
\begin{figure*}
\centering
\includegraphics[scale=1]{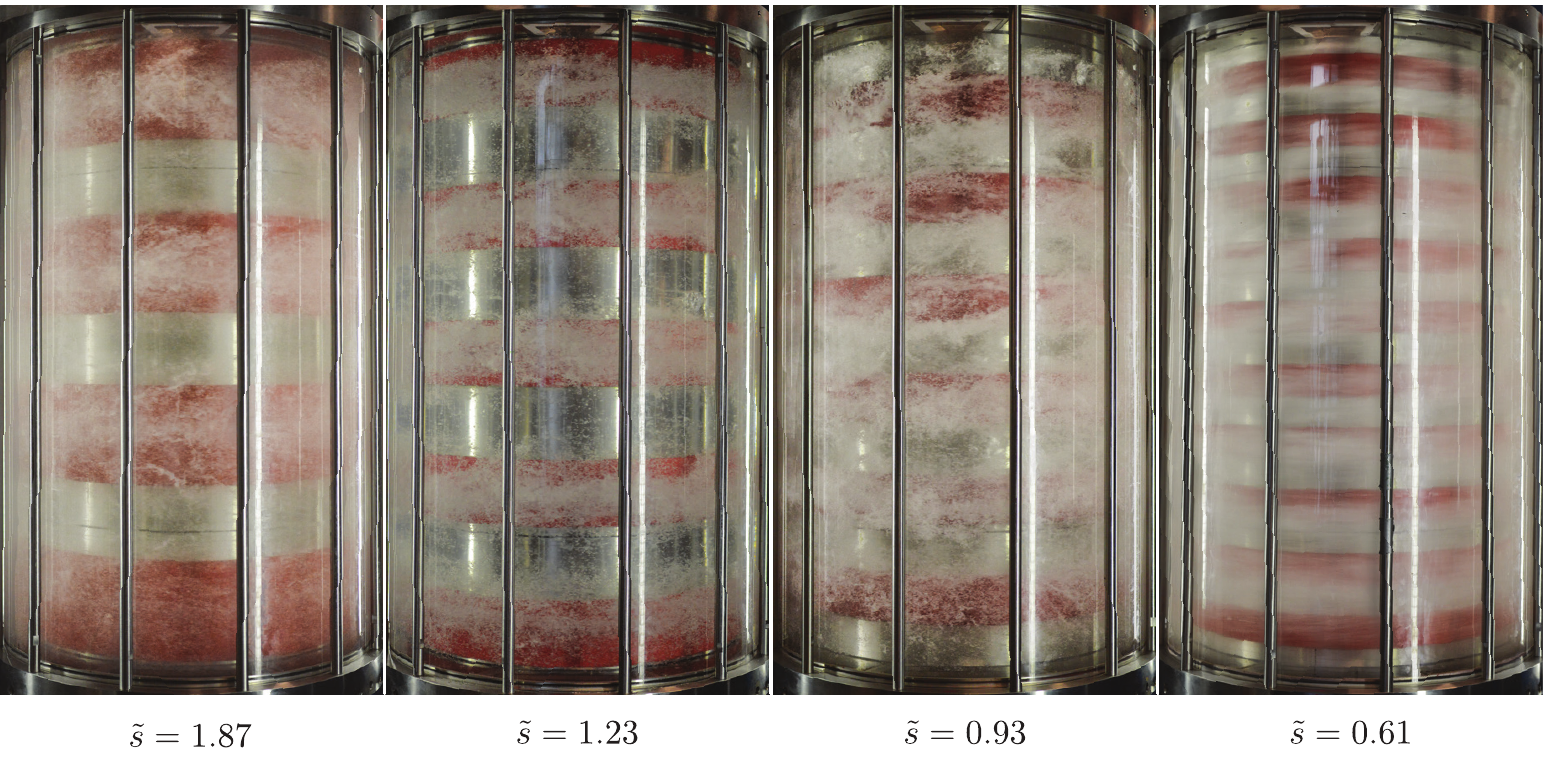}
\caption{Digitally enhanced photographs of the set~up, taken at $\Rey = 0.8\e{6}$ and $\alpha = \perc{1} $. From visual inspection it is clear that most bubbles reside at the rough patches, where the turbulent intensity of the flow is higher compared to the smooth patches. The effect is therefore two-fold: 1) compared to a completely smooth inner cylinder, for the same $\Rey$ there will be stronger turbulent mixing in the flow, resulting in a more even axial distribution of the air bubbles. And 2), at the rough-wall regions, the transfer of energy to the flow (or drag) is higher compared to the smooth-wall regions. The drag-reducing effect of bubbles is therefore at these positions of largest influence on the total drag. Hence, the bubbles move to the locations in the flow where they are needed most.}\label{fig:bubbles_photos}
\end{figure*}

Shown in figure~\ref{fig:bubbles_photos} are photographs of the experiment, taken at $\Rey = 0.8\e{6}$ with $\alpha = \perc{1}$ air in the working liquid. The bubbles show a preference to accumulate at the rough patches in the flow. This is best visible in the roughness configurations where the separation between the roughness bands is largest: $\tilde{s} = 1.87$ and $\tilde{s} = 1.23$. When the separation between roughness bands is smaller, in the $\tilde{s} = 0.93$ and $\tilde{s} = 0.61$ configuration, the bubbles can more easily travel between bands, leading to a more even axial bubble distribution.

\subsection{Secondary flow structure}
\begin{figure*}
    \includegraphics[scale=1]{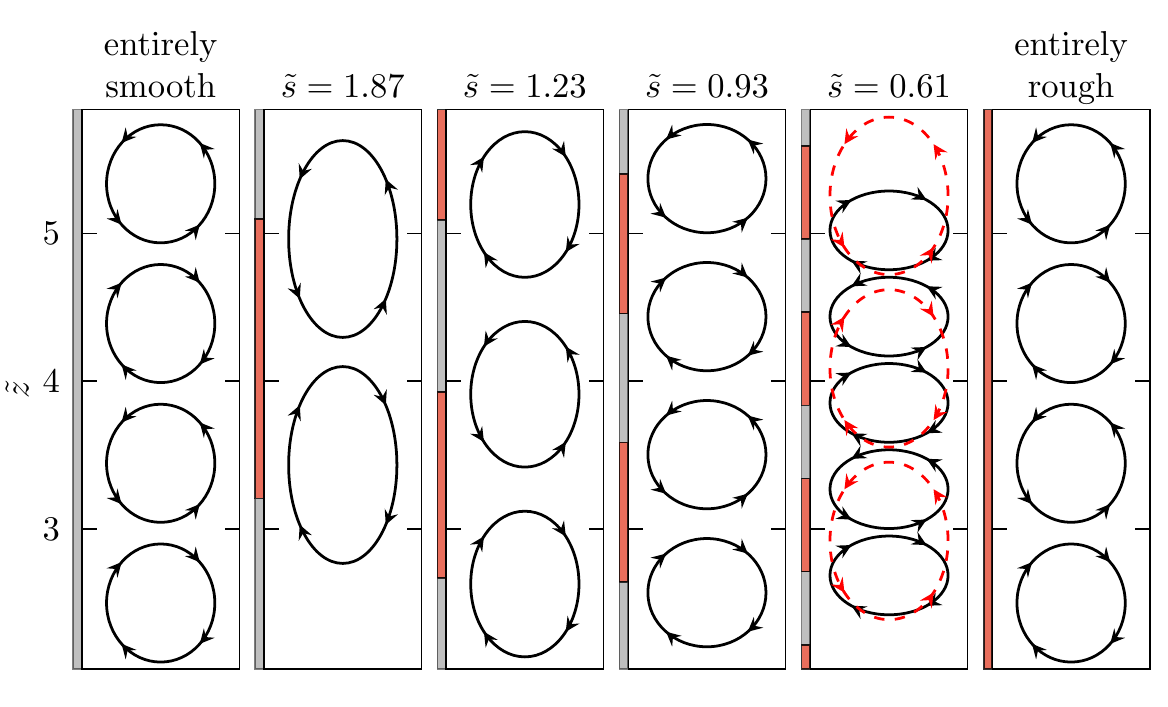}
    \caption{Position, size and rotational direction of the secondary flow structures, based on the inflow and outflow velocities on top of the smooth- and rough bands as reported by~\cite{Bakhuis2019} that were obtained using particle image velocimetry. The rolls are drawn for different roughness configurations $\tilde{s}$, in the gap between inner- (left) and outer cylinder (right). The height is normalized with the gap width $\tilde{z} = z/(r_o - r_i)$. The sandpaper roughness bands are indicated with red on the inner cylinder, the smooth bands are grey. For $\tilde{s}=0.61$ outflow regions from two adjacent roughness bands were seen to combine~\citep{Bakhuis2019}, hence, larger roll structures can also form and are indicated by the red-dashed arrows.}
    \label{fig:rolls}
\end{figure*}

To gain insight in the local flow organization, we refer to the work by~\cite{Bakhuis2019}. Based on their results from particle image velocimetry measurements, we draw in figure~\ref{fig:rolls} the positions and directions of the roll structures that are induced by the rough patches on the inner cylinder. The radially out- and inward flow at the rough- and smooth bands together form a roll~\citep{Bakhuis2019}. The locations of the rolls are determined by the boundaries between the rough and the smooth bands. For the configuration with the smallest bands, $\tilde{s} = 0.61$, the radially outward flows from two adjacent roughness bands are also seen to combine and form a larger roll~\citep{Bakhuis2019}. This can also be seen in figure~\ref{fig:rolls}, and may lead to variations in size and position of the secondary flows. Although for smooth walls it has been reported in literature that roll structures such as those that are observed for all different $\tilde{s}$ transport the air bubbles away from the inner cylinder by trapping them in their core~\citep{Mazzitelli2003,Climent2007,Lohse2018}, this is not what we observe here for the flow over rough surfaces. Instead, we find the bubbles in the radial outflow regions of the roughness, as that is where the turbulent intensity is highest.
This was shown by~\cite{Bakhuis2019}, who used Laser Doppler Anemometry to measure the velocities at mid-gap and different heights, to cover the flow above both smooth and rough bands. They found the standard deviation of the velocity to reach a peak at the centre of the rough bands, with a value of $\sigma(u_\theta)/u_i \approx 0.04$, where $u_i$ is the surface velocity of the inner cylinder. On the smooth bands, this had a value of $\sigma(u_\theta)/u_i \approx 0.03$. When the separation between roughness bands is too small, $\tilde{s} \leq 0.61$, this  phenomenon of velocity fluctuations following the underlying structure is lost~\citep{Bakhuis2019}. Also the rolls do no longer follow the topology of the roughness bands anymore, as is shown in figure~\ref{fig:rolls}.

We can use the data of~\cite{Bakhuis2019}, who measured $\sigma(u_\theta)/u_i$ at $\Rey = 0.8\e6$, to estimate a local Froude number $\Frcent$, which can quantify the preferred radial location of bubble accumulation. When we take $u_\theta$ in equation~\ref{eq:Fr_cent} to be $\approx u_i/2$ ~\citep{Huisman2013}, we can plug in the values reported by~\cite{Bakhuis2019}. For the bubble radius we refer to the work of \cite{vanGils2013}, who measured in the same setup the bubble diameter for smooth walls at $\Rey = 0.5\e6$ and $\Rey = 1.0\e6$, from which we approximate the bubble radius to be about \SI{0.5}{\mm} at $\Rey = 0.8\e6$. With the assumption that the bubble diameter is the same on the roughness as it is on the smooth surfaces, we find $\Frcent = 0.9$ above the smooth bands, and $\Frcent = 1.6$ above the rough bands. Hence, a \SI{75}{\percent} larger Froude number above the roughness, that may explain that we find more bubbles trapped at these locations.

\subsection{Torque measurements}

\begin{figure*}
  \includegraphics[scale=1]{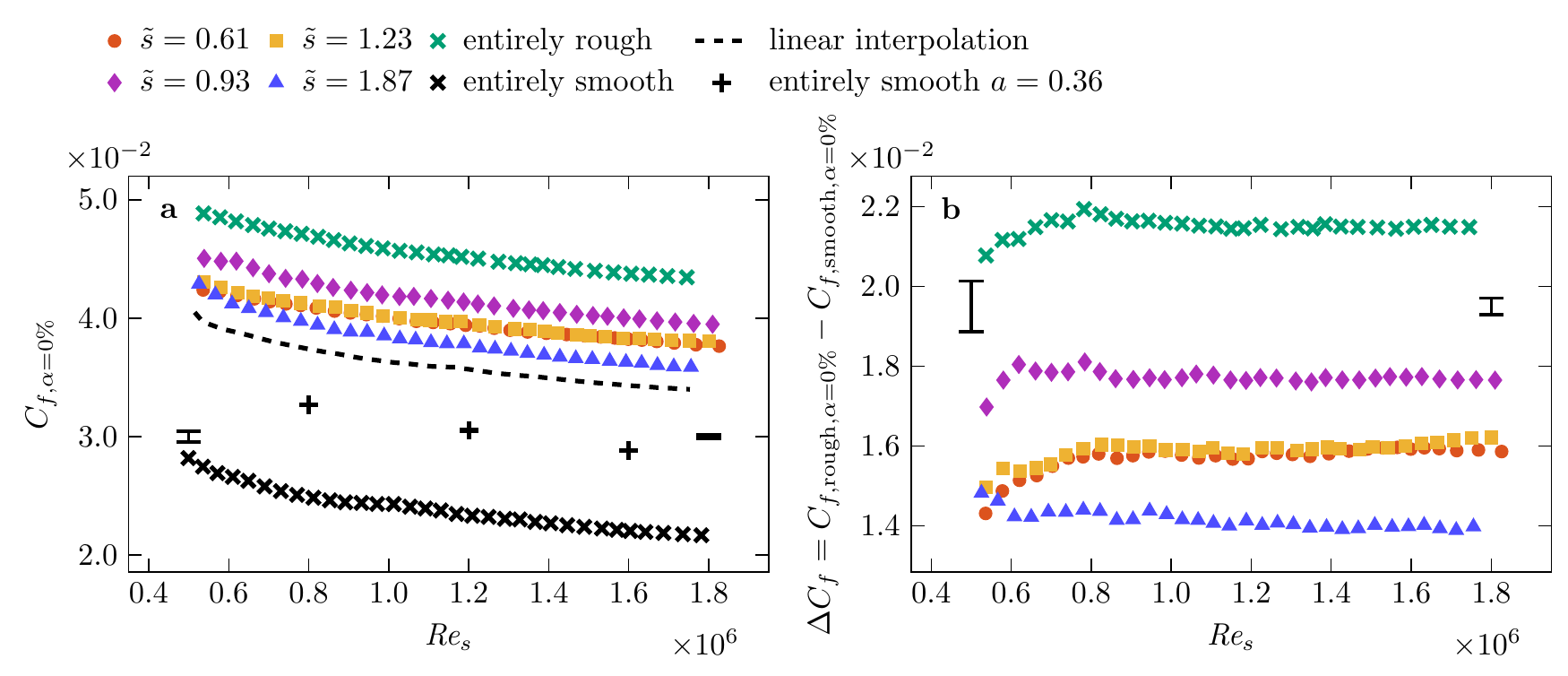}
  \caption{Results of torque measurements for $a=0$, plotted as skin friction coefficient $C_f$ versus (shear) Reynolds number $\Rey$ for a single phase flow with no air bubbles ($\alpha = \perc{0}$) in the working liquid. Shown in a) are the individual results. We also included a linear interpolation between the fully rough and fully smooth data included to arrive at a 56/44 rough/smooth distribution similar to the patched roughness data of certain $\tilde{s}$ (dashed line). Also included are the results of the counter-rotation measurements at $a = 0.36$, where the value of the torque is maximum (black pluses). Shown in b) are the differences in skin friction coefficient $\Delta C_f$ between the different rough and the smooth surface. Errors bars are shown in both graphs, based on the error in the torque sensor and measurement repeatability. The data for $\alpha = 0$ is the same as used in~\citep{Bakhuis2019}.}\label{fig:singlephase}
\end{figure*}

We first show the influence of the roughness in its different configurations on the skin friction coefficient $C_f$ versus $\Rey$ for single phase flow in figure~\ref{fig:singlephase}. A simple linear interpolation between the measured torque for a fully smooth and a fully rough inner cylinder to arrive at the same \perc{56} roughness coverage as for the partly rough surfaces with different $\tilde{s}$ underestimates the skin friction coefficient of those patchy rough surfaces. This is the result of secondary flow structures (rolls) that are created by the alternating rough and smooth bands. In figure~\ref{fig:singlephase}b, we plot the difference in skin friction coefficient between the rough surfaces and the smooth surfaces. For the fully rough surface, the skin friction coefficient is nearly doubled as compared to the smooth surface at the highest $\Rey$. We find the largest increase in skin friction coefficient for the roughness configuration $\tilde{s} = 0.93$. Based on our previous work~\citep{Bakhuis2019}, and other studies \citep{Chung2018}, this is explained as the roughness configuration in which the strongest roll structures are formed, as it is closest to $\tilde{s} = 1$. This is the most natural dimension for the roll, as its diameter matches the determining large length scale of the flow, which is in our case the width of the gap between the cylinders $d$. These rolls transport angular momentum from the inner cylinder (that drives the flow) to the outer cylinder. Hence, stronger rolls, can transport more angular momentum, which results in more torque and a larger skin friction coefficient. The same reasoning is used to explain that the configuration $\tilde{s} = 1.87$ will generate the least strong roll structures, and therefore gives the smallest increase in skin friction coefficient compared to the entirely smooth surface.

\begin{figure*}
  \includegraphics[scale=1]{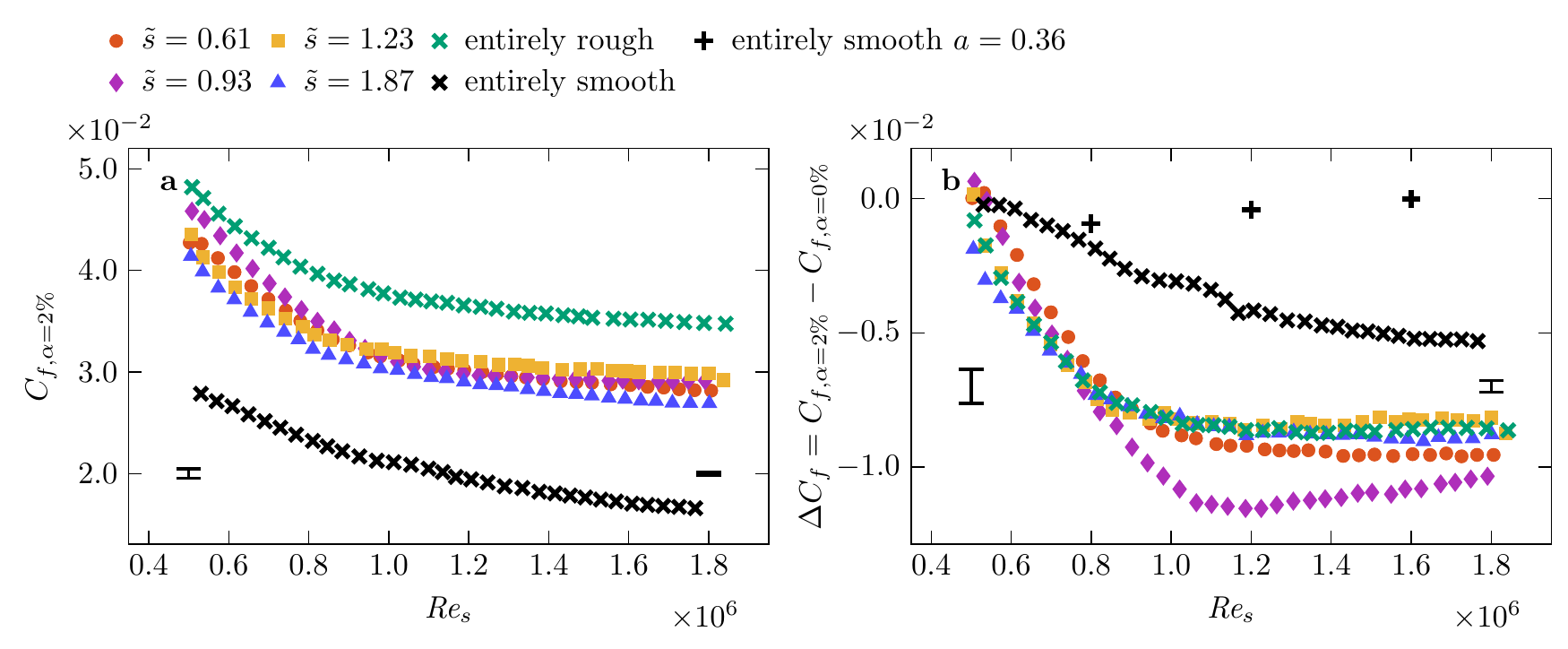}
  \caption{Results of torque measurements for $a=0$, plotted as skin friction coefficient $C_f$ versus shear Reynolds number $\Rey$ for a two-phase flow with 2 volume percent air bubbles ($\alpha = \perc{2}$) in the working liquid. Shown in a) are the individual results. Shown in b) are the differences in skin friction coefficient $\Delta C_f$ between the flow with $\alpha = \perc{2}$ and with $\alpha = \perc{0}$ over the same surface. Also included are the differences in $C_f$ for $\alpha = \perc{2}$ and $\alpha = \perc{0}$ of the counter-rotation measurements at $a = 0.36$ (black pluses). Errors bars are shown in both graphs, based on the error in the torque sensor and measurement repeatability. The data for $\alpha = 0$ is the same as used in~\citep{Bakhuis2019}.}\label{fig:two-phase}
\end{figure*}

 In figure~\ref{fig:two-phase}a we plot the skin friction coefficient versus $\Rey$ for the same roughness configurations as those in figure \ref{fig:singlephase}a, but now in the presence of bubbles. The working liquid contains 2 volume percent of air bubbles ($\alpha = \perc{2}$). Figure~\ref{fig:two-phase}b shows the difference in skin friction coefficient $\Delta C_f$ between a flow containing air bubbles ($\alpha=\perc{2}$) and a flow without air bubbles over the same roughness configuration. As a reference this is also shown for a fully smooth surface. Initially, up to $\Rey \approx 1.0\e{6}$, we find a very strong decrease in $C_f$ with $\Rey$ for the rough surfaces compared to the smooth surface when bubbles are introduced. We partly attribute this to increased levels of turbulence in the flow, due to the introduction of the roughness that leads to a more even axial distribution of the air bubbles at the lower shear Reynolds numbers. The other effect is that the bubbles prefer the regions of high turbulence intensity close to the roughness. This is also visible for the larger $\Rey > 1.0\e{6}$, where the introduction of bubbles leads to a greater decrease of the skin friction coefficient compared to the smooth surface. The $\Delta C_f$ does not significantly change with $\Rey$ for $\Rey > 1.0\e{6}$, as the axial distribution of air bubbles in the setup is well-mixed already, and does not change much with $\Rey$ anymore. Also the global structure of the secondary flows does not change over the course of the experiment with $\Rey$, and the roughness is always in the fully rough regime over the whole range of $\Rey$. Hence, there are no major changes in the relevant physical effects that determine the DR. It is clearly shown, that the drag reducing effect of the bubbles is stronger on rough surfaces compared to smooth surfaces, as is evident from the difference between $\Delta C_f$ on the fully rough and the fully smooth surface.

\subsection{Counter rotating outer cylinder}
\begin{figure*}
  \includegraphics[scale=1.1]{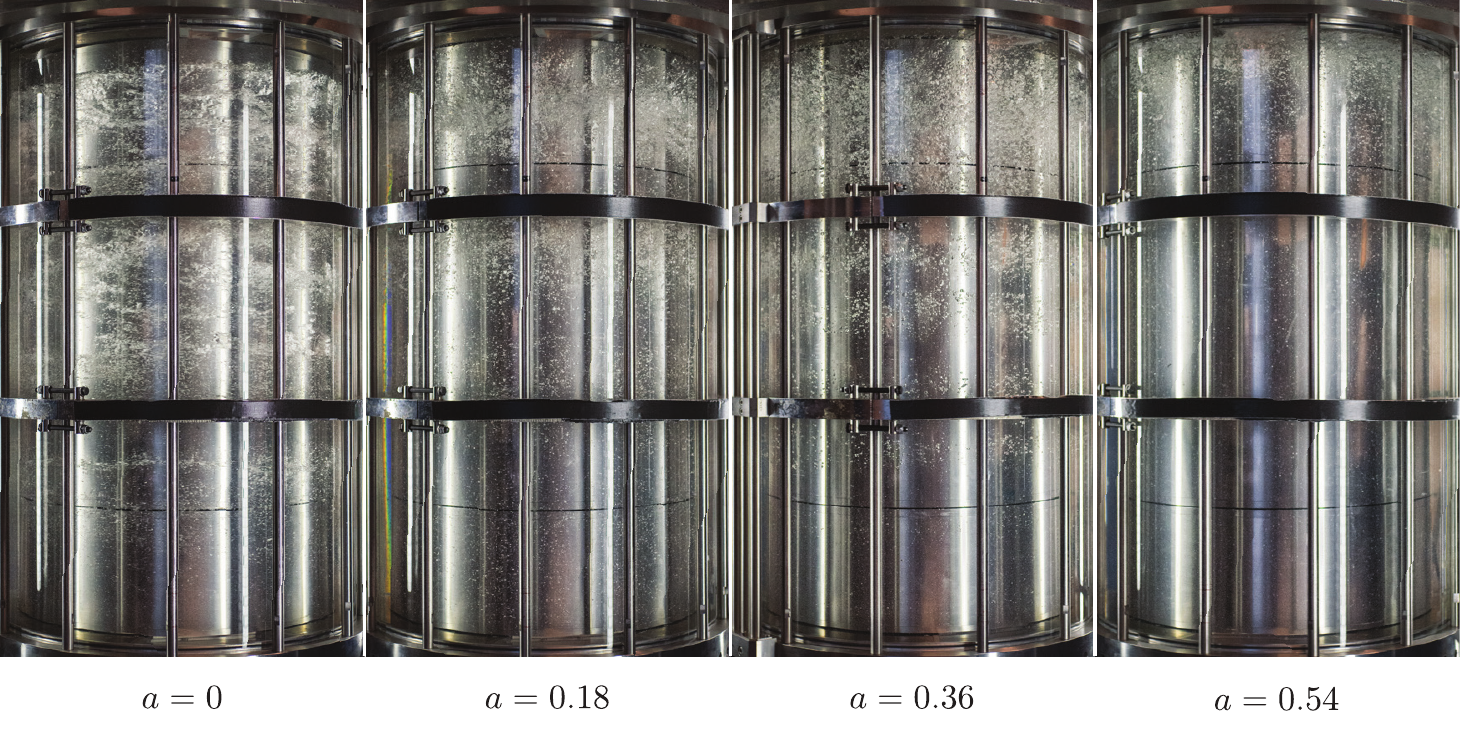}
  \caption{Digitally enhanced photographs of the set~up for the case of smooth walls, taken at $\Rey = 0.8\e{6}$ and $\alpha = \perc{1} $. The position of the air bubbles strongly depends on the ratio between rotation ratio $a = -\omega_o / \omega_i$. When $a = 0$, the mixing and axial distribution of bubbles is much more even compared to when $a > 0$. For $a>0$, stable roll structures are present in the flow that trap the bubbles, which is especially visible for the strongest roll structures at $a=0.36$.} \label{fig:bubbles_photos_different_a}
\end{figure*}

Shown in figure~\ref{fig:bubbles_photos_different_a} are snapshots of the flow with $\Rey = 0.8\e{6}$ and $\alpha = \perc{1}$ for different values of the rotation ratio $a$. For $a = 0$, no clear structure can be discovered in the flow. Although effects of buoyancy still play a role, the bubbles are more evenly distributed over the full height of the cylinders compared to the other $a > 0$ cases. When counter-rotation of the outer cylinder is introduced to the flow, turbulent Taylor-like vortices are formed in which the bubbles organise themselves. This is best observed for $a = 0.36$ when the rolls are strongest. When the rolls are weaker, for $a = 0.18$ and $a = 0.54$, the position of the bubbles is more affected by buoyancy~\citep{Spandan2018}. When the shear Reynolds number is larger, the rolls will be stronger and buoyancy effects smaller. Here we matched the shear Reynolds number between figure~\ref{fig:bubbles_photos} and figure~\ref{fig:bubbles_photos_different_a} to allow for a direct comparison. When these figures are compared, it also becomes evident that the roughness enhances mixing in the flow, resulting in a more even axial distribution of the bubbles. This illustrates the aforementioned trend of a stronger decrease in $C_f$ with $\Rey$ observed in figure~\ref{fig:two-phase}.

\begin{figure}
    \includegraphics[scale=1]{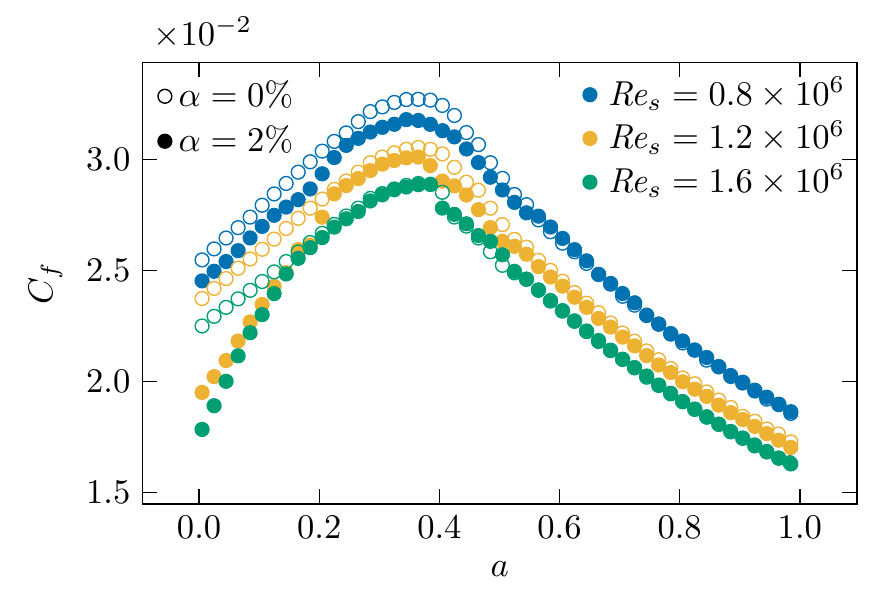}
    \caption{Results of the counter-rotation torque measurements, plotted as skin friction coefficient $C_f$ versus the rotation ratio $a = -\omega_o / \omega_i$ for three different Reynolds numbers for the case of smooth walls. Shown are the results of measurements without air bubbles $\alpha = \perc{0}$ and with two volume percent of air bubbles $\alpha = \perc{2}$ in the working liquid. These continuous measurement are done during \SI{48}{\minute} ($\Rey=0.8\times10^6$), \SI{72}{\minute} ($\Rey=1.2\times10^6$), and \SI{96}{\minute} ($\Rey=1.6\times10^6$). For the two highest Reynolds numbers, the rolls are so strong when $a>0.2$, that they transport the air bubbles away from the inner cylinder and the DR is lost. For the smallest Reynolds number are buoyancy effects still to strong to allow an even axial distribution of air, resulting in only minor DR.} \label{fig:asweep}
\end{figure}

The rolls capture the bubbles in their core and keep them away from the inner cylinder, which leaves them useless for drag reducing purposes. This is shown in figure~\ref{fig:asweep} where the skin friction coefficient $C_f$ is plotted versus $a$ for the three different Reynolds numbers for both $\alpha = \perc{2}$ and $\alpha = \perc{0}$. Whereas the configuration with the lowest $\Rey = 0.8\e{6}$ does show some DR up to $a = 0.6$, for the other two values of $\Rey$ the difference in $C_f$ is quickly reducing when $a > 0$. The differences between the values of $C_f$ at $\alpha = \perc{2}$ and $\alpha = \perc{0}$ at $a=0.36$ are included in figure~\ref{fig:two-phase}b.

When we look into the results of \cite{Dong2008}, and \cite{Huisman2014}, we expect that the turbulent fluctuations in the fluid reaches a maximum around $a \approx 0.36$. This is seen over the whole gap, but the increase is largest in the bulk region of the flow ($\approx \SI{55}{\percent}$), and smallest near the walls ($\approx \SI{25}{\percent}$)~\citep{Dong2008}. The centripetal Froude number $\Frcent$ will therefore also increase more in the bulk than near the walls, when counter-rotation is introduced, resulting in more trapping of bubbles in the bulk region.

\section{Discussion and conclusion}
The torque measurements presented in figure~\ref{fig:two-phase}b tell us that air bubble DR is more effective over sandpaper rough surfaces, compared to smooth surfaces. In particular when comparing this to our previous work, where we showed that riblet (obstacle) roughness on the inner cylinder of a Taylor--Couette set~up reduces air bubble DR~\citep{Verschoof2018}, it is more than obvious that the type of roughness is of large influence. For both types of roughness, secondary flows in the form of rolls are induced. However, for the sandpaper roughness the bubbles prefer to stay close to the roughness, instead of being carried away from the surface by the induced secondary flows. In the case of rib-like roughness, the induced rolls do carry the bubbles away from the surface~\citep{Verschoof2018}. When we induce rolls by introducing outer cylinder counter-rotation to a rotating smooth inner cylinder, we find almost no DR from the introduction of air bubbles to the flow. So in this case it is also the secondary flow structures that transport the air bubbles away from the surface.

As to why the bubbles prefer the regions in the flow near the rough patches, we argue that at these locations the turbulent intensity of the flow is largest. For heavy particles it is known that they accumulate in regions of minimum turbulent intensity, a phenomena known as turbophoresis~\citep{Reeks1983,Marchioli2002}. We drive the flow using the rotation of the inner cylinder (shear driven flow), and such the velocities will be \textit{larger} at the roughness bands where the bubbles cluster~\citep{Bakhuis2019}. This is in contrast to pressure driven flow configurations (such as pipe and channel flow), where the velocities will be \textit{lower} at the roughness. When air bubbles are thought of as light particles (compared to the working liquid) they move towards the regions of maximum turbulent intensity, where the local pressure will also be lowest~\citep{Mazzitelli2003,Climent2007,Loisy2017,Elghobashi2019,Mathai2019}, and hence end up in the high turbulent regions near the rough patches. As long as this effect is stronger than the effect of the rolls (secondary flows) which tend to move the bubbles away from the inner cylinder, we predict that DR will persist.

The influence of the turbulent intensity on the bubble position shows also up in the centripetal Froude number, that we used to explain the observed positions of the bubbles in flows with rough walls and flows with a counter rotating outer cylinder. Although we made several assumptions regarding the equal bubble size on rough and smooth surfaces, and we did not take into account the larger flow velocities found in the bulk of the flow above rough surfaces, the \SI{75}{\percent} larger $\Frcent$ for the rough surface is convincing. It could be expected that the bubbles are actually smaller above the rough surfaces, which might compensate for the velocity increase that is unaccounted for. Near the wall, the turbulent fluctuations will also be larger than at the mid-gap location where~\cite{Bakhuis2019} did their LDA measurements, also in accordance with~\cite{Huisman2013} and \cite{Berghout2019}.

\cite{Deutsch2004} found in their experiments with rough surfaces in a water tunnel the largest values of DR for the lowest flow velocities. We find for the rough surfaces in the low Reynolds number regime $\Rey < 1.0\e{6}$ a strongly increasing DR with $\Rey$, as can be seen in figure~\ref{fig:two-phase}b. This is attributed to the increased mixing effect with increasing Reynolds number that distributes the bubbles, versus a constant influence of gravity. This is also evident from the definition of the Froude number $\Fr$ in equation~\ref{eq:Fr}, since both $\Rey$ and $\Fr$ scale linearly with the rotation rate of the inner cylinder $\omega_i$. This can therefore be attributed as an effect related to our flow geometry. In the regime of large Reynolds numbers, $\Rey > 1.2\e{6}$, where mixing dominates over gravity, we find the DR no longer changes with $\Rey$. Although the study of~\cite{Deutsch2004} is more oriented towards the influence of air bubble injection rate and different surface roughness heights $k$, this is a surprising difference.

We have shown that bubble DR is more effective on rough surfaces. However, a careful examination of figure~\ref{fig:singlephase}~and~\ref{fig:two-phase} reveals that the skin friction coefficient of a smooth surface without air bubbles present in the flow ($\alpha = \perc{0}$), is still lower than that of a rough surface with air bubbles ($\alpha = \perc{2}$). For industrial applications of air bubble DR in the maritime industry, this means that it remains very important to keep the hull of a ship clean and smooth. An important side note here is that in these kind of practical application, a perfect hydrodynamically smooth surface --- like we used in this research --- is almost never encountered, since the cost of such a surface finish is too large for these applications. These surfaces will therefore always feature some roughness at a relevant length scale making the flow hydrodynamically rough. Therefore it is very relevant to realize that for such rough surfaces, although the principles and mechanisms we have learned and observed from experiments using smooth surfaces are very similar, the roughness does have an influence on the bubble position and the resulting DR, and should not be neglected.

\begin{acknowledgments}
We would like to thank Gert-Wim Bruggert, Martin Bos and Bas Benschop for their continuous technical support over the years with the T$^3$C. We acknowledge stimulation discussions with Pieter Berghout on roughness. We thank Dominic Tai and You-An Lee for their help in the lab. This research is supported by the project “GasDrive: Minimizing emissions and energy losses at sea with LNG combined prime movers, underwater exhausts and nano hull material” (project 14504) of the Netherlands Organisation for Scientific Research (NWO), domain Applied and Engineering Sciences (TTW). D.B. and C.S. acknowledge financial support from VIDI grant No. 13477, and the Natural Science Foundation of China under grant nos. 91852202 and 11672156. We also acknowledge financial supported by NWO-I and the ERC under the Advanced Grant ``Physics of liquid-vapor phase transition'' and from MCEC.
\end{acknowledgments}

\bibliography{Draft_roughness_bubbles}

\end{document}